\documentstyle[preprint,revtex,eqsecnum]{aps}

\begin{document}
\preprint{PITT 92-07}

\vspace{1mm}
\preprint{CMU-HEP-92-25}
\draft
\begin{title}
{\bf Phase Transitions out of Equilibrium:}\\
{\bf Domain Formation and Growth}
\end{title}
\author{{\bf Daniel Boyanovsky, Da- Shin Lee}}
\begin{instit}
Department of Physics and Astronomy\\ University of
Pittsburgh \\
Pittsburgh, P. A. 15260, U.S.A.
\end{instit}
\author{\bf Anupam Singh}
\begin{instit}
Department of Physics\\Carnegie Mellon University\\
Pittsburgh, P.A. 15213, U.S.A.
\end{instit}
\begin{abstract}
We study the dynamics of phase transitions out of
equilibrium in  weakly coupled scalar field theories. We
consider the case in which there is a rapid supercooling
from an initial  symmetric phase in  thermal equilibrium at temperature
$T_i>T_c$ to a final state at low temperature $T_f \approx
0$. In particular we study the formation and growth of
correlated domains out of equilibrium. It is shown that the
dynamics of the process of domain formation and growth
(spinodal decomposition) cannot be studied in perturbation theory, and
a non-perturbative
self-consistent Hartree approximation is used to study the
long time evolution. We find in weakly coupled theories that
the size of domains grow at long times
 as $\xi_D(t) \approx \sqrt{t\xi(0)}$.
The size of the domains and the amplitude of the fluctuations grow up to a
 maximum time $t_s$ which in weakly coupled theories is estimated to be

\[ t_s \approx -\xi(0)\ln\left[\left(\frac{3\lambda}{4\pi^3}\right)^
{\frac{1}{2}}\left(\frac{(\frac{T_i}{2 T_c})^3}{[\frac{T^2_i}{T^2_c}-
1]}\right)\right] \]

with $\xi(0)$ the zero temperature
correlation length. For very weakly coupled theories, their
final size is several times the zero temperature correlation length. For
strongly coupled theories the final size of the domains is comparable to
the zero temperature correlation length and the transition proceeds faster.
\end{abstract}

\section{\bf Introduction and Motivation:}

Phase transitions play a fundamental role in our understanding of
the interplay between cosmology and particle physics in extreme
environments. It is widely
accepted that many different phase transitions took place in the early
universe at different energy (temperature) scales and with remarkable
consequences at low temperatures and energies. In particular broken
symmetries, and possibly the observed baryon asymmetry in the universe.

Phase transitions are an essential ingredient in inflationary models of
the early universe\cite{abbott,guth1,guth2,linde1,steinhardt,linde2}.
The importance of the description of phase transitions in
extreme environments was recognized long time ago and efforts were
devoted to their description in relativistic
quantum field theory at finite temperature\cite{kirzhnits,dolan,sweinberg}.
 For a very thorough account of  phase transitions in the early universe
 see the reviews by Brandenberger\cite{brandenberger},
 Kolb and Turner\cite{kolb} and Linde\cite{linde}.

 The methods used to study the {\it equilibrium} properties of
 phase transitions are by now well understood and widely used,
 in particular field theory at finite temperature and effective
 potentials\cite{kapusta}.

 These methods, however, are restricted to a {\it static} description
 of the consequences of the phase transition, but can hardly be used to
 understand the {\it dynamics} of the processes involved  during
 the phase transition. In particular, for example, the effective potential
 that is widely used to determine the nature of a phase transition and
 static quantities like critical temperatures etc, is {\it irrelevant}
 for the description of the dynamics. The effective potential corresponds
 to (minus) the {\it equilibrium} free energy density as a function of the
 order parameter. This is a static quantity, calculated in equilibrium, and
 in particular to one loop order it is complex within the region of
 homogeneous field
 configurations in which $V''(\phi)< 0$, where $V(\phi)$ is the classical
 potential. This was already recognized in the early treatments by Dolan
 and Jackiw\cite{dolan}.

 In statistical mechanics, this region is referred to as the
 ``spinodal'' and corresponds to a sequence of states which are
  thermodynamically unstable.

  At zero temperature, the imaginary part of
  the effective potential has been identified with the decay rate of this
  particular unstable state\cite{weinbergwu}.

The use of the static effective potential to describe the dynamics of
phase transitions has been criticized by many authors, among them
Mazenko, Unruh and
Wald\cite{mazenko}. These authors argued that phase transitions in
typical theories will occur  via the formation and growth of
correlated domains inside which the field will relax to the value of the
minimum of the equilibrium free energy fairly quickly. It is now believed
that this picture may be correct for {\it strongly coupled theories} but
is not accurate for weak couplings.

The mechanism that is responsible for a typical second order
phase transition from a initially symmetric high temperature state is
fairly well known. When the temperature becomes lower than the critical
temperature, long-wavelength fluctuations become unstable and
begin to grow and the field becomes correlated inside ``domains'', the order
parameter (the expectation value of the volume average of the field),
remaining zero all throughout the transition.

Recently, de Vega and one of the authors, have studied
the influence of these instabilities in the evolution of the order
parameter out of equilibrium\cite{boyveg} in the case when a non-zero
(but small) initial value of the order parameter was assumed.

In this work we continue the study of the {\it non-equilibrium}
aspects of second
order phase transitions in typical scalar field theories, with a
view towards a deeper understanding of the {\it dynamics} of
phase transitions in inflationary scenarios of the early universe.
In particular trying to describe the process of domain formation and growth
in the case in which the initial state is symmetric and in equilibrium at
a temperature higher than the critical temperature and cooled down below
the critical temperature.

Although there have been several attempts to study the time evolution
of the scalar field either in flat spacetime or de Sitter
space\cite{ringwald,kripfganz,leutwyler,eboli,samiulla}, to our
knowledge there has not as yet been a consistent treatment of the
{\it dynamics} of domain formation and growth.

Recently, Kolb and Wang\cite{wang}
have reported on an  equilibrium study of the  static properties of
domains produced in late-time phase transitions, but it becomes a pressing
issue to understand  the process of domain formation
and growth, especially the time scales involved and the size of the
correlations.

Although our initial motivation, and ultimate goal, is to study the
dynamics of the phase transition in an expanding universe, in this
article we
report our studies on the dynamics of the phase transitions in Minkowski
space. We do not attempt in
this article to study the non-equilibrium properties in the necessarily more
complicated setting of inflationary cosmologies, and restrict ourselves to
introducing the methods of non-equilibrium quantum statistical mechanics
and apply them to the study of formation and growth of domains in flat
spacetime.

We would like to point out at this stage, that the situation under
consideration is very different from the classical description of the
process of spinodal decomposition in statistical mechanics. The classical
approach to spinodal decomposition is based on a ``coarse grained'' time
dependent Landau-Ginzburg equation, which is first order in time and purely
dissipative. Thermal fluctuations are usually introduced as a Langevin
noise,typically uncorrelated\cite{langer,guntonmiguel},
that obeys the fluctuation dissipation relation.

In our case we are studying a {\it quantum field theory}, the Heisenberg
field equations are second order, non-dissipative (in Minkowski space),
 and both quantum and
thermal fluctuations are present in the initial state (density matrix),
furthermore, as is typical in these scalar theories, the order parameter is
not conserved.

This article is organized as follows: in  section 2 we present our
arguments that suggest that phase transitions in expanding cosmologies must
be studied away from equilibrium for weakly coupled theories. We emphasize
that the important long-wavelength modes that become unstable below
the critical temperature and whose dynamics is relevant for the process of
phase separation and domain growth will easily be out of equilibrium during
the transition for weakly coupled theories.

In section 3, we introduce our model and the methods of non-equilibrium
statistical mechanics as applied to the description of the dynamics of the
phase transition.

In section 4, we analyze the real time correlation functions in zeroth
order and obtain the first quantitative expressions, a scaling law for
the size of the domains and the growth of the amplitude of the fluctuations.
In section 5, we carry out a perturbative calculation and
show quantitatively that the dynamics cannot be studied within a
perturbative framework.

In section 6 we introduce a non-perturbative self consistent
Hartree approximation to study
the evolution of correlations and growth of domains. We provide an analytic
and numerical analysis of the process of domain growth and establish a
scaling law for the size of the domains at long times, and
an estimate for the maximum size of the domains for very weakly coupled
theories.

We summarize our findings and pose further questions in section 7.

\section{\bf The case for a non-equilibrium description:}

Before entering into the technical details, let us sketch
the arguments that
suggest that when the temperature is  very near the critical
temperature,
the relevant dynamics {\it must} be studied out of
equilibrium.

As in any dynamical process, in order to try to describe the
time evolution
of the system, one must first try to determine the typical
time scales
involved in the different dynamical processes. This
understanding becomes
more pressing when one tries to understand the dynamics of
phase transitions.

In particular, is it possible to describe the phase
transition in an
environment in which the temperature is changing at some
particular rate,
in local thermodynamic equilibrium?. To address this issue
one must compare
the typical collisional relaxational rates of the particles,
to the rate
of change of the temperature.

The typical collisional
relaxation rate for a process at energy E in the heat bath
is given by
$\Gamma(E) = \tau^{-1}(E) \approx n(E)\sigma(E)v(E)$ with
$n(E)$ the number
density of particles  with this energy E, $\sigma(E)$ the
scattering cross
section at this energy and $v(E)$ the velocity of the
incident beam of
particles. The lowest order (Born approximation) scattering
cross
section in a typical
scalar theory with quartic interaction is $\sigma(E) \approx
\lambda^2 / E^2$.
At very high temperatures, $T \gg m_{\Phi}$, with $m_{\Phi}$
the mass of the
field, $n(T) \approx T^3$, the internal energy density is
$U/V \approx T^4$,
and the average energy per particle is $\langle E \rangle
\approx T$, and
$v(T) \approx 1$.
Thus the typical collisional relaxation rate is
$\Gamma(T) \approx \lambda^2 T$. In an expanding universe,
the conditions for
local equilibrium will prevail provided that $\Gamma(T) \gg
(\dot{a(t)}/a(t))=H(t)$,
with $a(t)$ the FRW scale factor. If this is the case, the
collisions occur
very quickly compared to the expansion rate, and particles will
equilibrate. This
argument applies to the collisional relaxation of high
frequency
(short-wavelength) modes for which $k \gg m_{\Phi}$. This
may be understood as follows.
Each ``external leg'' in the scattering process
considered, carries  typical momentum and energy $k,E
\approx T > T_c $.
But in these typical theories $T_c \approx m_{\Phi}/
\sqrt{\lambda}$.  Thus
for weakly coupled theories $T_c \gg m_{\Phi}$.

To obtain an order of magnitude estimate, we concentrate
near the phase transition at $T \approx T_c \approx 10^{14}
{\mbox{Gev}}$, $H \approx T^2 / M_{Pl} \approx 10^{-5}T$,
this implies that for $\lambda > 0.01$ the conditions for
local equilibrium may prevail. However, in weakly coupled
inflaton models of
inflation, phenomenologically the coupling is bound by the
spectrum of density fluctuations to be $\lambda \approx
10^{-12}-10^{-14}$\cite{linde,weakcoup}. Thus, in these weakly
coupled theories  the
conditions for local equilibrium of high energy modes may
not be achieved. One may, however, assume that although the
scalar field
is weakly self-coupled, it has strong coupling to the heat
bath (presumably
other fields in the theory) and thus remains in local
thermodynamic equilibrium.

This argument, however, applies to the collisional
relaxation of {\it short
wavelength modes}. We observe, however, that these type of
arguments are
{\it not valid} for the dynamics of the {\it long-wavelength
modes} at
temperatures {\it below} $T_c$, for the following reason.

At very high temperatures, and in local equilibrium, the
system is in the
disordered phase with $\langle\Phi\rangle = 0$ and {\it
short ranged
correlations}, as measured
by the equal time correlation function (properly subtracted)
\begin{eqnarray}
\langle\Phi(\vec{r},t)\Phi(\vec{0},t)\rangle & \approx & T^2
\exp[-
|\vec{r}|/ \xi(T)]
\label{correlationfunction} \\
\xi(T)                           & \approx &
\frac{1}{\sqrt{\lambda} T}
\approx \xi(0)(\frac{T_c}{T}) \label{hitcorrelationlength}
\end{eqnarray}

As the temperature drops near the critical temperature, and
below, the phase
transition occurs. The onset of the phase transition is
characterized by the
instabilities of long-wavelength fluctuations, and the
ensuing growth of
correlations. The field begins to correlate over larger
distances, and correlated domains will form and grow.
 If the
initial value of the order parameter is zero, it will remain
zero throughout the transition.
This is the
process of spinodal decomposition, or phase separation. This
growth of correlations, cannot be described as a process in
local thermodynamic equilibrium.

These instabilities are
manifest in the {\it equilibrium} free energy in the form of
imaginary parts, and the equilibrium free energy is not a
relevant quantity to study the {\it dynamics}.

These long
wavelength modes whose instabilities trigger the phase
transitions have very
slow dynamics. This is the phenomenon of critical slowing
down that is
observed experimentally in binary mixtures and numerically
in typical
simulations of phase transitions. The long wavelength
fluctuations correspond to coherent collective behavior in
which degrees of freedom become correlated over large
distances. These collective long-wavelength modes have
extremely
slow relaxation near the phase transition, and they do
not have many available low energy decay channels.
Certainly through the phase transition, high frequency,
short wavelength modes
may still remain in local equilibrium by the arguments
presented above (if the coupling is sufficiently strong),
they
have many channels for decay and thus will maintain local
equilibrium through the phase transition.

To make this argument more quantitative, consider the
situation in which the final temperature is below the
critical value and early times after the transition. For
small amplitude fluctuations of the field, long wavelength
modes ``see'' an inverted harmonic oscillator and the
amplitude fluctuations begin to grow as (see below)
\begin{eqnarray}
\langle\Phi_{k}(t)\Phi_{-\vec{k}}(t)\rangle & \approx &\exp[2
W(k)t]
\label{unstable} \\
W(k)                      &  =      & \sqrt{\mu^2(T)-
\vec{k}^2}
\label{bigomega} \\
\mu^2(T)                  &   =     & \mu^2(0)[1-
(\frac{T_f}{T_c})^2]
\end{eqnarray}
for $\vec{k}^2 < \mu^2(T)$.

In particular this situation, modelled with the ``inverted harmonic
oscillators'' is precisely the situation thoroughly and clearly studied by
Guth and Pi\cite{guthpi} and Weinberg and Wu\cite{weinbergwu}.

The time scales that must be compared for the dynamics of
these
instabilities,
are now the growth rate $ \Gamma(k) \approx \sqrt{\mu^2(T)-
\vec{k}^2}$ and
the expansion rate $H \approx T^2/M_{Pl} \approx (10^{-
5})T$, if the
expansion rate is comparable to the growth rate, then the
long wavelength
modes that are unstable and begin to correlate may be in
local thermodynamic
equilibrium through the cooling down process. Using $ T_c
\approx \mu(0) /
\sqrt{\lambda}$, we must compare $ [1-(\frac{T}{T_c})^2]^{\frac{1}{2}}$ to
$10^{-5}/ \sqrt{\lambda}$. Clearly for weakly coupled theories, or
very near the
critical temperature, the growth rate of the
unstable modes will be much slower than the rate of cooling
down, and the
phase transition will be supercooled, similarly to a
``quenching process''
from a high temperature, disordered phase to a supercooled
low temperature  situation. For example, for $\lambda \approx
10^{-12}$  the
growth rate of long wavelength fluctuations, is much smaller than the
expansion rate even for a final temperature $T_f \approx 0$, and the
long wavelength modes will be strongly supercooled.

As mentioned previously, we do not attempt in this article
to study the situation in an expanding gravitational
background, and limit ourselves to studying the dynamics of
the phase transition in Minkowski space by modelling the
important features that are relevant for the phase
transition in a weakly coupled theory.  Our goals here are to
introduce the methods to study this type of phase transitions out of
equilibrium, and to study the physics of domain formation and growth
within a simplified situation. Eventually we propose to extend these
methods to the case of an inflationary background.

We do not envisage here to account for the cooling down
process (which
requires a clear understanding of gravitational effects, and time scales)
and restrict
ourselves to {\it assuming} a supercooled phase transition
in a weakly
coupled theory and propose a particular model to
understand this situation in Minkowski space.

\section{\bf Statistical Mechanics out of equilibrium:}

A period of rapid temperature change, may be modeled by
considering a time dependent Hamiltonian with a {\it time
dependent mass term} of the form

\begin{eqnarray}
  H(t) & = & \int_{\Omega} d^3x \left\{
\frac{1}{2}\Pi^2(x)+\frac{1}{2}(\vec{\nabla}\Phi(x))^2+\frac
{1}{2}m^2(t)
\Phi^2(x)+\frac{\lambda}{4!}\Phi^4(x) \right \}
\label{timedepham} \\
m^2(t) & = & m^2_i \Theta(-t) - m^2_f \Theta(t)
\label{massoft} \\
m^2_i  & = & \mu^2[\frac{T^2_i}{T^2_c}-1]
\label{massinitial} \\
m^2_f  & = &  \mu^2[1-\frac{T^2_f}{T^2_c}] \label{massfinal}
\end{eqnarray}
with $ \mu^2 > 0 \; ; \; T_i > T_c \; ; \; T_f \ll T_c$.
The introduction of the $T_i \; ; \; T_f$ in the above mass
term, is just a {\it parametrization} of the model, which
incorporates the
ingredients of a high temperature state at $t<0$ and a low
temperature
situation for $t>0$. Again, the mechanism that drives the
phase transition
may either be a period of rapid inflation or a sudden
coupling to a heat
bath at a much lower temperature. The above parametrization
incorporates
by hand this ``rapid supercooling'' situation.

Clearly this is a simplification, but in view of the above
comments, we believe that this approximation is justified
insofar as we are trying to understand the dynamics of the
instabilities of the long-wavelength modes and the growth of
correlations in weakly coupled theories.
This assumption of a rapid ``quench'' may be
relaxed at the expense of complications. As it will become
clear below, this approximation will allow us to obtain
analytic results and to perform explicit calculations.
Furthermore, we assume that for all times $t <0$ the system
is in thermal equilibrium at the initial temperature $T_i$,
thus described by the density matrix
\begin{eqnarray}
\hat{\rho}_i  & = & e^{-\beta_i H_i} \label{initialdesmat}\\
          H_i & = & H(t<0) \label{initialham}
\end{eqnarray}

In the Schroedinger picture, the density matrix evolves in
time as
\begin{equation}
\hat{\rho(t)} = U(t)\hat{\rho}_iU^{-1}(t) \label{timedesmat}
\end{equation}
with $U(t)$ the time evolution operator.

An alternative, and equally valid interpretation is that we
consider an initial condition in which the system is in
thermodynamic equilibrium at temperature $T_i \gg T_c$ in
the symmetric phase for $t<0$, and evolved in time with a
Hamiltonian with a ``negative mass squared'' that allows for
broken symmetry  states for $t>0$.

This interpretation  in fact describes the situation studied by Guth
and Pi\cite{guthpi} and Weinberg and Wu\cite{weinbergwu}. These authors
prepare an initial gaussian state or density matrix, and study the time
evolution of this initially prepared state with a Hamiltonian for a
collection of ``inverted harmonic oscillators''. Preparing an initial state,
and evolving it with a Hamiltonian of which the initial state is {\it not
an eigenstate} (in the language of density matrices, the density matrix
does not commute with the Hamiltonian) is the quantum mechanics equivalent
of a ``quenching process'' or a ``sudden approximation''. It is in this
sense that we are thus generalizing the situation studied by the above
authors.

The expectation
value of any operator is thus
\begin{equation}
\langle {\cal{O}} \rangle(t) = Tr e^{-\beta_i H_i} U^{-1}(t)
{\cal{O}}U(t)/ Tr e^{-
\beta_i H_i}
\label{expecvalue}
\end{equation}
This expression may be written in a more illuminating form
by choosing an arbitrary time $T <0$ for which
$U(T) = \exp[-iTH_i]$ then we may write
$\exp[-\beta_i H_i] = \exp[-iH_i(T-i\beta_i -T)] = U(T-
i\beta_i,T)$.
Inserting in the trace $U^{-1}(T)U(T)=1$,
commuting $U^{-1}(T)$ with $\hat{\rho}_i$ and using the
composition property of the evolution operator, we may write
(\ref{expecvalue}) as
\begin{equation}
\langle {\cal{O}}\rangle(t) = Tr U(T-i\beta_i,t) {\cal{O}}
U(t,T)/ Tr
U(T-i\beta_i,T) \label{trace}
\end{equation}
The numerator of the above expression has a simple meaning:
start at time $T<0$, evolve to time $t$, insert the operator
$\cal{O}$ and evolve backwards in time from $t$ to $T<0$,
and along the negative imaginary axis from $T$ to $T-
i\beta_i$.
The denominator, just evolves along the
negative imaginary axis from $T$ to $T-i\beta_i$. The
contour in
the numerator may be extended to an arbitrary large positive
time
$T'$ by inserting $U(t,T')U(T',t)=1$ to the left of
$\cal{O}$ in
(\ref{trace}), thus becoming
\begin{equation}
\langle {\cal{O}}\rangle(t) = Tr U(T-
i\beta_i,T)U(T,T')U(T',t){\cal{O}}U(t,T)
/Tr U(T-i\beta_i,T)
\end{equation}
The numerator now represents the process of evolving from
$T<0$
to $t$, inserting the operator $\cal{O}$, evolving further
to
$T'$, and backwards from $T'$ to $T$ and down the negative
imaginary axis to $T-i\beta_i$. This process is depicted in
the
contour of figure (1). Eventually we take $T \rightarrow -
\infty
\; ; \; T' \rightarrow \infty$. It is straightforward to
generalize
to  real time correlation functions of Heisenberg picture
operators.

This formalism allows us also to study the general case in
which both the mass and the coupling depend on time. The
insertion of the operator ${\cal{O}}$, may be achieved as
usual by introducing currents and taking variational
derivatives with respect to them.

Because the time evolution operators have
the interaction terms in them, and we would like to generate
a perturbative expansion and Feynman diagrams, it is
convenient to introduce source terms for {\it all} the time
evolution operators in the above trace. Thus we are led to
consider the following generating functional
\begin{equation}
Z[J^+, J^-, J^{\beta}]= Tr U(T-i\beta_i,T;J^{\beta})
U(T,T';J^-)U(T',T;J^+)\label{generatingfunctional}
\end{equation}
The denominator in (\ref{trace}) is simply $Z[0,0,0]$ and
may be obtained in a series expansion in the interaction by
considering $Z[0,0,J^{\beta}]$.
By inserting a complete set of field eigenstates between the
time evolution operators, finally, the generating functional
$Z[J^+,J^-,J^{\beta}]$ may be written as
\begin{eqnarray}
Z[J^+,J^-,J^{\beta}] & = & \int D \Phi D \Phi_1 D \Phi_2
\int {\cal{D}}\Phi^+ {\cal{D}}\Phi^-
{\cal{D}}\Phi^{\beta}e^{i\int_T^{T'}\left\{{\cal{L}}[\Phi^+,
J^+]-
{\cal{L}}[\Phi^-,J^-]\right\}}\times   \nonumber\\
                     &   &  e^{i\int_T^{T-
i\beta_i}{\cal{L}}[\Phi^{\beta}, J^{\beta}]}
\label{pathint}
\end{eqnarray}
with the boundary conditions $\Phi^+(T)=\Phi^{\beta}(T-
i\beta_i)=\Phi \;
; \; \Phi^+(T')=\Phi^-(T')=\Phi_2 \; ; \; \Phi^-
(T)=\Phi^{\beta}(T)=
\Phi_1$.
This may be recognized as a path integral along the contour
in complex time shown in Figure (1).
As usual the path integrals over the quadratic forms may be
evaluated and one obtains the final result for the partition
function
\begin{eqnarray}
& & Z[J^+,J^-, J^{\beta}] = \exp{\left\{i\int_{T}^{T'}dt
\left[{\cal{L}}_{int}(-i\delta/\delta J^+)-
{\cal{L}}_{int}(i\delta/\delta
J^-)\right] \right \}} \times \nonumber \\
& &
\exp{\left\{i\int_{T}^{T-
i\beta_i}dt{\cal{L}}_{int}(-i\delta/\delta J^{\beta})
\right\}} \exp{\left\{\frac{i}{2}\int_c
dt_1\int_c dt_2 J_c(t_1)J_c(t_2)G_c (t_1,t_2) \right\}}
 \label{partitionfunction}
\end{eqnarray}

Where $J_c$ stands for the currents on the contour as shown
in figure (1), $G_c$ are the Green's functions on the
contour\cite{niemisemenoff}, and again,
the spatial arguments were suppressed.

In the limit $T \rightarrow \infty$, the contributions from
the terms in which one of the currents is $J^+$ or $J^-$ and
the other is a $J^{\beta}$ vanish when computing correlation
functions in which the external legs are at finite {\it real
time}, as a consequence of the Riemann-Lebesgue lemma. For
this {\it real time} correlation functions, there is no
contribution from the $J^{\beta}$ terms that cancel between
numerator and denominator. Finite temperature enters through
the boundary conditions on the Green's functions (see
below). For the calculation of finite {\it real time}
correlation functions, the generating functional simplifies
to\cite{calzetta,calzettahu}
\begin{eqnarray}
Z[J^+,J^-] & = & \exp{\left\{i\int_{T}^{T'}dt\left[
{\cal{L}}_{int}(-i\delta/\delta J^+)-
{\cal{L}}_{int}(i\delta/\delta
J^-)\right] \right \}} \times \nonumber \\
           &   & \exp{\left\{\frac{i}{2}\int_T^{T'}
dt_1\int_T^{T'} dt_2 J_a(t_1)J_b(t_2)G_{ab} (t_1,t_2)
\right\}}
 \label{generatingfunction}
\end{eqnarray}
with $a,b = +,-$.

  The Green's functions that enter in the integrals along
the  contours in equations (\ref{partitionfunction},
\ref{generatingfunction})
  are given by (see above references)
\begin{eqnarray}
G^{++}(\vec{r}_1,t_1;\vec{r}_2,t_2)  & = &
G^{>}(\vec{r}_1,t_1;\vec{r}_2,t_2)\Theta(t_1-t_2) +
G^{<}(\vec{r}_1,t_1;\vec{r}_2,t_2)\Theta(t_2-t_1)
\label{timeordered}\\
G^{--}(\vec{r}_1,t_1;\vec{r}_2,t_2)  & = &
G^{>}(\vec{r}_1,t_1;\vec{r}_2,t_2)\Theta(t_2-t_1) +
G^{<}(\vec{r}_1,t_1;\vec{r}_2,t_2)\Theta(t_1-t_2)
\label{antitimeordered} \\
G^{+-}(\vec{r}_1,t_1;\vec{r}_2,t_2)  & = & -
G^{<}(\vec{r}_1,t_1;\vec{r}_2,t_2) \label{plusminus}\\
G^{-+}(\vec{r}_1,t_1;\vec{r}_2,t_2)  & = & -
G^{>}(\vec{r}_1,t_1;\vec{r}_2,t_2) = -
G^{<}(\vec{r}_1,t_1;\vec{r}_2,t_2)
\label{minusplus}\\
G^{>}(\vec{r}_1,t_1;\vec{r}_2,t_2)   & = &
\langle\Phi(\vec{r}_1,t_1)\Phi(\vec{r}_2,t_2)\rangle
\label{greater} \\
G^{<}(\vec{r}_1,T;\vec{r}_2,t_2)     & = &
G^{>}(\vec{r}_1,T-i\beta_i;\vec{r}_2,t_2)
\label{periodicity}
\end{eqnarray}

The condition(\ref{periodicity}) is recognized as the
periodicity condition in imaginary time and is a result of
considering an equilibrium situation for $t<0$. The
functions $G^{>}, \; G^{<}$ are homogeneous solutions of the
quadratic form, with appropriate boundary conditions and
will be constructed explicitly below.

This formulation in terms of time evolution along a contour in complex
time has been used many times in non-equilibrium statistical mechanics.
To our knowledge the first to introduce this formulation were
Schwinger\cite{schwinger} and Keldysh\cite{keldysh} (for an early account
see Mills\cite{mills}). There are many clear articles in the literature
using this techniques to study real time correlation
functions\cite{calzetta,calzettahu,jordan,landsman,semenoffweiss,kobeskowalski,paz}.

Our goal is to study the formation and growth of domains and
the time evolution of the correlation functions. In
particular, the relevant quantity of interest is the {\it equal time}
correlation function
\begin{eqnarray}
S(\vec{r};t) & = &
\langle\Phi(\vec{r},t)\Phi(\vec{0},t)\rangle
\label{equaltimecorr} \\
S(\vec{r};t) & = & \int \frac{d^3 k}{(2\pi)^3}e^{i\vec{k}
\cdot \vec{r}} S(\vec{k};t) \label{strucfac} \\
S(\vec{k};t) & = & \langle\Phi_{\vec{k}}(t)\Phi_{-
\vec{k}}(t)\rangle =
(-iG^{++}_{\vec{k}}(t;t))\label{strufack}
\end{eqnarray}
where we have performed the Fourier transform in the spatial
coordinates (there still is spatial translational and
rotational invariance). Notice that at equal times, all the
Green's functions are equal, and we may compute any of them.

Clearly in an equilibrium situation this equal time correlation function
will be time independent, and will only measure the
{\it static correlations}. In the present case, however, there is a non
trivial time evolution arising from the departure from equilibrium of the
initial state. This correlation function will measure the correlations in
space, and their time dependence.

The function $G^{>}_{\vec{k}}(t,t')$ is constructed from the
homogeneous solutions to the operator of quadratic
fluctuations
\begin{eqnarray}
\left[\frac{d^2}{dt^2} + \vec{k}^2 +
m^2(t)\right]{\cal{U}}_k^{\pm} & = & 0 \label{homogeneous}
 \end{eqnarray}
with $m^2(t)$ given by (\ref{massoft}).

The boundary conditions on the homogeneous solutions are
\begin{eqnarray}
{\cal{U}}_k^{\pm}(t<0) & = & e^{\mp i
\omega_{<}(k)t}\label{boundaryconditions} \\
\omega_{<}(k)          & = &
\left[\vec{k}^2+m^2_i\right]^{\frac{1}{2}}\label{omegaminus}
\end{eqnarray}
corresponding to positive frequency (particles) and negative
frequency (antiparticles) (${\cal{U}}_k^{+}(t);
{\cal{U}}_k^{-}(t)$, respectively).

The solutions are as follows:
i): stable modes ($\vec{k}^2 > m^2_f$)
\begin{eqnarray}
{\cal{U}^+}_k(t) & = & e^{-i\omega_{<}(k) t}\Theta(-t)+
\left(a_k e^{-i\omega_{>}(k)t}+b_k
e^{i\omega_{>}(k)t}\right)\Theta(t) \label{stablemodes} \\
{\cal{U}^-}_k(t) & = & \left({\cal{U}^+}_k(t)\right)^{*} \\
\omega_{>}(k)    & = & \sqrt{\vec{k}^2-m^2_f}
\label{omegaplus} \\
a_k              & = &
\frac{1}{2}\left(1+\frac{\omega_{<}(k)}{\omega_{>}(k)}
\right) \\
b_k              & = & \frac{1}{2}\left(1-
\frac{\omega_{<}(k)}{\omega_{>}(k)}\right) \\
\end{eqnarray}

ii): unstable modes ($\vec{k}^2 < m^2_f$)
\begin{eqnarray}
{\cal{U}^+}_k(t) & = & e^{-i\omega_{<}(k) t}\Theta(-t)+
\left(A_k e^{W(k)t}+B_k e^{-W(k)t}\right)\Theta(t)
\label{unstablemodes} \\
{\cal{U}^-}_k(t) & = & \left({\cal{U}^+}_k(t)\right)^{*} \\
W(k)             & = & \sqrt{m^2_f-\vec{k}^2} \label{bigW}
\\
A_k              & = & \frac{1}{2}\left(1-
i\frac{\omega_{<}(k)}{W(k)}\right) \; \; ; \; \; B_k =
(A_k)^*  \\
\end{eqnarray}
With these mode functions, and the periodicity condition
(\ref{periodicity}) we find
\begin{eqnarray}
G^{>}_k(t,t') & = & \frac{i}{2\omega_< (k)} \frac{1}{1-
e^{-
\beta_i \omega_< (k)}}\left[{\cal{U}}_k^+(t)
{\cal{U}}_k^-
(t')+ e^{-\beta_i\omega_< (k)} {\cal{U}}_k^-(t)
{\cal{U}}_k^+(t') \right] \label{finalgreenfunc} \\
G^{<}_k(t,t') & = & G^{>}(t',t)
\end{eqnarray}

 The zeroth order equal time Green's function becomes
\begin{equation}
 (G^{>}_{\vec{k}}(t;t))= \frac{i}{2 \omega_{<}(k)}
 \coth[\beta_i\omega_{<}(k)/2]
\end{equation}
for $t< 0$, and

\begin{eqnarray}
(G^{>}_{\vec{k}}(t;t)) & = & \frac{i}{2 \omega_{<}(k)} \{
\left[1+2A_kB_k\left[ \cosh(2W(k)t)-1 \right]\right]
 \Theta(m^2_f-\vec{k}^2)
\nonumber  \\
                       & + &
\left[1+2a_kb_k\left[ \cos(2 \omega_{>}(k)t)-1
\right]\right]
 \Theta(\vec{k}^2-m^2_f)\}
\coth[\beta_i\omega_{<}(k)/2]
\end{eqnarray}
for $t > 0$.

The first term, the contribution of the unstable modes,
reflects the growth of correlations because of the
instabilities and will be the dominant term at long times.

\section{\bf Zeroth order correlations:}

Before proceeding to study the correlations in higher orders
in the coupling
constant, it will prove to be very illuminating to
understand the behavior
of the equal time non-equilibrium correlation functions at
tree-level.
Because we are interested in the  growth of correlations, we
will study only
the contributions of the unstable modes.

 The integral of the equal time correlation function over
all wave
vectors shows the familiar short distance divergences.  From
the above
expression; however, it is clear that these may be removed
by subtracting (and also multiplicatively renormalizing)
this correlation function at $t=0$.
The contribution of the stable modes to the
subtracted and multiplicatively renormalized correlation function is
always bounded in time and
thus uninteresting
for the purpose of understanding the growth of the
fluctuations.

We are thus
led to study {\it only} the contributions of the unstable
modes to the  subtracted and renormalized
correlation function, this contribution is finite and unambiguous.

For this purpose it is convenient to introduce the following
dimensionless quantities
\begin{equation}
\kappa = \frac{k}{m_f}  \; \; ; \; \;  L^2 = \frac{m_i^2}{m_f^2}=
\frac{\left[T^2_i-T^2_c\right]}{\left[T^2_c-T^2_f\right]}
\; \; ; \; \; \tau = m_f t  \; \; ; \; \;  \vec{x} = m_f \vec{r}
\label{dimensionless}
\end{equation}
Furthermore for the unstable modes $\vec{k}^2 < m^2_f$, and
for initial
temperatures  larger than the critical temperature
$T^2_c = 24 \mu^2 / \lambda$,  we can approximate $\coth[\beta_i
\omega_{<}(k)/2] \approx
2 T_i / \omega_{<}(k)$. Then, at tree-level, the
contribution of the
unstable modes to the subtracted structure factor
(\ref{strufack})
$S^{(0)}(k,t)-
S^{(0)}(k,0)=(1/m_f){\cal{S}}^{(0)}(\kappa,\tau)$ becomes
\begin{eqnarray}
{\cal{S}}^{(0)}(\kappa,\tau)   & = &
\left( \frac{24}{\lambda [1-\frac{T^2_f}{T^2_c}]} \right)^{\frac{1}{2}}
\left(\frac{T_i}{T_c} \right)
\frac{1}{2\omega^2_{\kappa}} \left( 1+
\frac{\omega^2_{\kappa}}{W^2_{\kappa}}
\right) \left[ \cosh(2W_{\kappa}\tau)-1 \right]
\label{strucuns} \\
              \omega^2_{\kappa}& = & \kappa^2+L^2 \label{smallomegak} \\
  W_{\kappa}                   & = & 1-\kappa^2 \label{bigomegak}
\end{eqnarray}

To obtain a better idea of the growth of correlations, it is
convenient to
introduce the scaled correlation function
\begin{equation}
{\cal{D}}(x,\tau) = \frac{\lambda}{6m^2_f}\int^{m_f}_0
\frac{k^2 dk}{2\pi^2}\frac{\sin(kr)}{(kr)}[S(k,t)-S(k,0)]
\label{integral}
\end{equation}
The reason for this is that the minimum of the tree level
potential occurs
at $\lambda \Phi^2 /6 m^2_f =1$, and the inflexion
(spinodal) point,
at $\lambda \Phi^2 /2 m^2_f =1$, so that ${\cal{D}}(0,\tau)$
measures the
excursion of the fluctuations to the spinodal point and
beyond as the correlations grow in time.

At large $\tau$ (large times),
the  product $\kappa^2 {\cal{S}}(\kappa,\tau)$ in
(\ref{integral}) has a very sharp peak at
$\kappa_s = 1/ \sqrt{\tau}$. In the region $x < \sqrt{\tau}$
the integral
may be done by the  saddle point approximation and we obtain for
$T_f/T_c \approx 0$ the large time behavior
\begin{eqnarray}
{\cal{D}}(x,\tau) & \approx & {\cal{D}}(0,\tau)
\exp[-\frac{x^2}{8\tau}]
\frac{\sin(x/ \sqrt{\tau})}{(x/ \sqrt{\tau})}
\label{strucfacxtau} \\
{\cal{D}}(0,\tau) & \approx & \left(\frac{\lambda}{12
\pi^3}\right)^
{\frac{1}{2}}\left(\frac{(\frac{T_i}{2 T_c})^3}{[
\frac{T^2_i}{T^2_c}-
1]}\right)\frac{\exp[2\tau]}{\tau^{\frac{3}{2}}}
\label{strucfactau}
\end{eqnarray}

 Restoring dimensions, and recalling that the zero
temperature correlation
 length is $\xi(0) = 1/\sqrt{2} \mu$,
 we find that for $T_f \approx 0$ the amplitude of the
fluctuation inside a
 ``domain'' $\langle \Phi^2(t)\rangle$, and the ``size'' of
a  domain  $\xi_D(t)$ grow as
 \begin{eqnarray}
 \langle \Phi^2(t)\rangle & \approx &  \frac{\exp[\sqrt{2}t/
\xi(0)]}
 {(\sqrt{2}t/ \xi(0))^{\frac{3}{2}}} \label{domainamplitude}
\\
 \xi_D(t)      & \approx & (8\sqrt{2})^{\frac{1}{2}}
 \xi(0)\sqrt{\frac{t}{\xi(0)}}
\label{domainsize}
 \end{eqnarray}

 An important time scale corresponds to the time $\tau_s$ at
which the  fluctuations
 of the field sample beyond the spinodal point. Roughly
speaking when this
 happens, the instabilities should shut-off as the mean
square root
 fluctuation of the field $\sqrt{\langle\Phi^2(t)\rangle}$
is now probing the stable  region.
 This will be seen explicitly below when we study the
evolution  non-perturbatively in the Hartree approximation and
the fluctuations are incorporated self-consistently in the evolution
equations.
 In zeroth order we estimate this time
from the condition
 $3{\cal{D}}(0,t) = 1$, we use
 $\lambda= 10^{-12} \; ; \;  T_i/T_c=2$,
 as representative parameters
 (this value
 of the initial temperature does not have  any particular physical
 meaning and was chosen only as  representative). We find
 \begin{equation}
 \tau_s \approx 10.15 \label{spinodaltime}
 \end{equation}
 or in units of the zero temperature correlation length
 \begin{equation}
 t \approx 14.2 \xi(0)
 \end{equation}
 for other values of the parameter $\tau_s$ is found from
the above condition
 on (\ref{strucfactau}).

 These  are some of the main results of this work.

 \section{\bf Perturbation theory and its demise:}

 The results presented in the previous section, rely on a
zero-order (tree
 level) analysis of the non-equilibrium correlation
function. Clearly
 one needs to incorporate the effects of the interaction.
 The non-equilibrium
 formalism introduced above lends itself to a diagrammatic
expansion of the
 non-equilibrium correlation functions. We now present a
one-loop calculation
 of the equal time correlation function
 $\langle\Phi_{\vec{k}}(t)\Phi_{-\vec{k}}(t)\rangle$.

 There are two vertices,
 corresponding to forward ($+$) and backward ($-$) time
propagation, with
 opposite couplings and four different propagators as given
in equations
 (\ref{timeordered}-\ref{minusplus}) (see figure (2.a)). The
Feynman rules
 are the standard ones. The Feynman diagrams that contribute
up to one loop
  to the structure factor (\ref{strufack})

 \[\langle\Phi_{\vec{k}}^+(t)\Phi_{-\vec{k}}^-(t)\rangle =
\langle\Phi_{\vec{k}}(t)
 \Phi_{-\vec{k}}(t)\rangle = iG^{+-}_{k}(t,t) \]

 are depicted in figure (2.b). Then up to one loop, and in
terms of the zero
 order Green's functions, we find
 \begin{eqnarray}
 \langle\Phi_{\vec{k}}(t)\Phi_{-\vec{k}}(t)\rangle  & = &
(-
iG^{<}_k(t,t)) \nonumber \\
                                        & + &
                     \frac{\lambda}{2}\int_{-\infty}^{t}dt_1
  \int \frac{d^3q}{(2\pi)^3}  \{ G^{>}_{0,q}(t_1,t_1)
  \left[(G^{>}_{0,k}(t,t_1))^2-
(G^{<}_{0,k}(t,t_1))^2\right]\times \nonumber \\
                                        &   & \coth[\beta_i
\omega_{<}(q)/2]
 \coth[\beta_i \omega_{<}(k)/2] \} \label{oneloop}
\end{eqnarray}
where in the one loop integral we wrote the finite
temperature Green's
functions in terms of the zero temperature ones
$G^{>}_{0,q}$. Clearly
because of the complicated time dependence, the Dyson's
series for the
propagator may not be summed exactly and we only analyze
here the one loop
contribution given above. Before proceeding further with the
analysis, let
us understand  the renormalizations
that are necessary.

It becomes more illuminating to write the one-loop
contribution explicitly
in term of the mode functions,

\begin{eqnarray}
\langle\Phi_{\vec{k}}^+(t)\Phi_{-\vec{k}}^-(t)\rangle & = &
\frac{{\cal{U}}^+_k(t){\cal{U}}^-_k(t)}{2\omega_{<}(k)}
\coth[\beta_i \omega_{<}(k)/2]+ \nonumber \\
                         &   &\frac{\lambda}{2}\int_{-\infty}^t dt_1\int
\frac{d^3q}{(2\pi)^3}\left(\frac{i}{2\omega_{<}(q)}\right)
\left(\frac{i}{2\omega_{<}(k)}\right)^2
{\cal{U}}^+_q(t_1){\cal{U}}^-_q(t_1)
\times \nonumber \\
                         &   &
\left[ \left({\cal{U}}^+_k(t){\cal{U}}^-_k(t_1)\right)^2-
       \left({\cal{U}}^+_k(t_1){\cal{U}}^-_k(t)\right)^2
\right]\times \nonumber \\
                         &   & \coth[\beta_i
\omega_{<}(q)/2]
 \coth[\beta_i \omega_{<}(k)/2] \label{onelup}
\end{eqnarray}
Clearly, in the one-loop contribution, the terms with
wavevectors $k$
correspond to the ``external legs'', whereas the terms with
$q$ which are
integrated over, correspond to the loop line in figure (2.b)
(because at equal time  the $++$ and $--$ terms are equal).

First let us study the above contribution for $t<0$, in
which case we
should recover the usual result.
In this case both $t; \; t_1 <0$, and performing
the time integral with an adiabatic cutoff we find
\begin{eqnarray}
\langle\Phi_{\vec{k}}(t)\Phi_{-\vec{k}}(t)\rangle     & = &
\frac{1}{2\omega_{<}(k)}\coth[\beta_i \omega_{<}(k)/2] -
\frac{\lambda}{2}\int
\frac{d^3q}{(2\pi)^3}\left(\frac{1}{2\omega_{<}(q)}\right)
\left(\frac{1}{2\omega_{<}(k)}\right)^3 \times \nonumber \\
                         &   &    \coth[\beta_i
\omega_{<}(q)/2]
 \coth[\beta_i \omega_{<}(k)/2] \label{tlesszero}
\end{eqnarray}

In fact this is the familiar equal time Green's function up
to one loop
 of the time independent theory. The one loop term has
temperature
 independent  ultraviolet
 divergent contribution arising from the q-integral.
Introducing an upper
 momentum
cutoff $\Lambda$ and an arbitrary renormalization scale
${\cal{K}}$ we obtain
\begin{equation}
I_{div}(t_1 < 0) =
\int
\frac{d^3q}{(2\pi)^3}\left(\frac{1}{2\omega_{<}(q)}\right)=
\frac{1}{8\pi^2}\left[{\Lambda}^2-
m^2_i\ln\left(\frac{\Lambda}{{\cal{K}}}
\right)\right] \label{divertlesszero}
\end{equation}

In any case, it becomes clear that the potential divergences
of the one-loop
contribution, can be traced to the integral (again only the
zero temperature
contribution is divergent)
\begin{equation}
I_{div}(t_1)= \int
\frac{d^3q}{(2\pi)^3}\left(\frac{1}{2\omega_{<}(q)}\right)
        {\cal{U}}^+_q(t_1){\cal{U}}^-_q(t_1)\label{diverg}
\end{equation}

For $t>0$, the time integral in the one-loop correction can
be split into
the integral $\int^0_{-\infty}dt_1$ and $\int^t_0dt_1$. In
the first integral
(from $-\infty$ to $0$), ${\cal{U}}^+_q(t_1){\cal{U}}^-
_q(t_1)=1$ and
the divergence
structure is the same as that analyzed for $t<0$. In the
second integral
(from $0$ to $t$)) the divergent contribution arises solely
from the
{\it stable modes} as the unstable modes are cutoff at
$q=m_f$.  By analyzing
the product of the mode functions, it becomes clear that the
time dependent
part ($2a_qb_q \cos[2\omega_{>}(q)t_1]$) will yield to a
finite contribution
because the strong oscillations (for $t_1 > 0$) ensure
convergence at large
momenta. Thus the divergent term arises only from the time
independent
term,  ($a^2_q+b^2_q$), in the product of mode functions.
Finally we find
the divergent term to be temperature independent and
for $t_1 > 0$  given by
\begin{equation}
I_{div}=\frac{1}{4\pi^2}
\int_{m_f}^{\infty}dq\left(\frac{q^2}
{\sqrt{q^2+m_i^2}}\right)\left\{1+\frac{1}{2}\left(\frac{m_i
^2+m_f^2}
{q^2-m^2_f}\right)\right\} \label{divint}
\end{equation}
Again in terms of an ultraviolet cutoff ($\Lambda$)  and renormalization
scale (${\cal{K}}$), we find
\begin{equation}
I_{div}(t_1 > 0)=
\frac{1}{8\pi^2}\left[{\Lambda}^2-(-
m^2_f)\ln\left(\frac{\Lambda}{{\cal{K}}}
\right)\right] \label{divertgreatzero}
\end{equation}
These divergences may be cancelled by introducing a local
but {\bf time
dependent} counterterm in the original Lagrangian density
\begin{eqnarray}
{\cal{L}}_{ct}  & = & \delta m^2(t)\Phi^2(\vec{r},t)
\label{counterlagrangian} \\
\delta m^2(t)   & = & -\frac{\lambda}{8\pi^2}\int
dq\frac{q^2}
{\sqrt{q^2+m_i^2}}\left\{\Theta(-t)+\Theta(t)\Theta(q^2-
m^2_f)(a^2_q+b^2_q)
\right\} \label{counterterm}
\end{eqnarray}
On the forward and backward time contour for the non-
equilibrium theory,
this counterterm translates in the two counterterm insertions
shown in figure
(2.c).  The introduction of these counterterms renders
finite
the one-loop contribution to all the non-equilibrium Green's
functions as may  now be easily checked.

Having disposed of the renormalization problem, we must
however address the
issue of the instabilities. The instabilities and growth of
correlations at
zeroth-order had been analyzed before. We now realize that
in the loop
integral there is a contribution to the loop from the
integration over the
unstable modes which will enhance the exponential growth in
the correlation  functions.

The maximum instability in the one-loop term is when the
mode functions for
both momenta $q  ; k$ are unstable, ($q^2  ;  k^2 < m^2_f$).
For
the initial temperature $T_i > T_c$, for these values
of the momenta we use
the high temperature approximation $\coth[\beta_i \omega_{<}/2] \approx
2T_i/ \omega_{<}$. It is convenient to introduce the
dimensionless quantities defined in equation
(\ref{dimensionless}) and $Q=q/m_f$. Using equations
(\ref{massinitial},\ref{massfinal})
and $T_c^2= 24 \mu^2/\lambda$, and the same conventions as
for the zeroth order structure factor  given by equations
(\ref{smallomegak},\ref{bigomegak},\ref{dimensionless}),
we obtain for the {\it most unstable contribution} to the
one-loop correction to
the structure factor $S^{(1)}(k,t) =
(1/m_f){\cal{S}}^{(1)}(\kappa,\tau)$
\begin{eqnarray}
{\cal{S}}^{(1)}(\kappa, \tau) & = &
\left(-\frac{6}{\pi^2}\right)\frac{T^2_i}{T^2_c[1-
(T^2_f/T^2_c)]}\int^{\tau}_{0}
d\tau_1\int_0^1 dQ
\frac{Q^2}{\omega^2_Q\omega^2_{\kappa}W_{\kappa}}\times
\nonumber \\
                              &   & \sinh[W_{\kappa}(\tau-
\tau_1)]
\{1+\frac{1}{2}\left(1+\frac{\omega^2_{Q}}{W^2_{Q}}\right)
[\cosh(2W_{Q}\tau_1)-1]\} \times \nonumber \\
                              &   &
\left[\left(1+\frac{\omega^2_{\kappa}}{W^2_{\kappa}}\right)\
\cosh[W_{\kappa}
(\tau+\tau_1)]+\left(1-
\frac{\omega^2_{\kappa}}{W^2_{\kappa}}\right)
\cosh[W_{\kappa}(\tau-\tau_1)]\right] \label{s1unst}
\end{eqnarray}
The integral over $\tau_1$ may be carried out yielding a
rather cumbersome
result, but it becomes clear that this result will grow
roughly as the square
of the zeroth order result at large $\tau$. Introducing the
scaled correlation
function as in equation (\ref{integral}) both for the zeroth order and the
one-loop order
${\cal{D}}^{(0)}(x,\tau)\; ; \; {\cal{D}}^{(1)}(x,\tau)$
respectively, in figure
(3.a) we show the behavior for $3{\cal{D}}^{(0)}(0,\tau)$
(solid lines) and
$3{\cal{D}}^{(1)}(0,\tau)$ (dashed lines) for the values of
the parameters
$\lambda = 10^{-12} \; , \; (T_i/T_c)=2$. It is clear that
eventually the
one loop term becomes {\bf much larger} than the tree level
term even in
the case of very weak coupling. This is a
consequence of the instabilities and the growth of correlations that are a
hallmark of the phase transition. Clearly the {\it dynamics} of the phase
transition {\it cannot be studied in perturbation theory}. In fact this
result in a  very quantitative manner confirms the ideas that the onset
of the transition and the time evolution of the system after the phase
transition cannot be  studied perturbatively.

\section{\bf Beyond perturbation theory: Hartree approximation}

It became clear from the analysis of the previous section
that perturbation
theory is inadequate to describe the non-equilibrium
dynamics of the phase
transition, precisely because of the instabilities and the
growth of
correlations. This growth is manifest in the Green's
functions that enter in
any perturbative expansion thus invalidating any
perturbative approach. Higher
order corrections will have terms that grow exponentially
and faster than the
previous term in the expansion. And even for very weakly
coupled theories,
the higher order corrections eventually become of the same
order as the lower order terms.

As the correlations and fluctuations grow, field
configurations start sampling the stable region beyond the spinodal point.
This will result in  a slow down in the
growth of correlations, and eventually  the unstable growth
will shut-off.
When this happens, the state may be described by correlated domains
with equal
probabibility for both phases inside the domains. The
expectation value of
the field in this configuration will be zero, but inside
each domain, the
field will acquire a value very close to the value in
equilibrium at the
minimum of the {\it effective potential}. The size of the
domain in this
picture will depend on the time during which correlations
had grown enough so that
fluctuations start sampling beyond the spinodal point.

Since this physical picture may not be studied within
perturbation theory,
we now introduce a {\it non-perturbative} method based on a
self-consistent Hartree approximation\cite{conjecture,lawrie,chang}.

The self-consistent Hartree approximation is implemented as
follows:
in the initial Lagrangian write
\begin{equation}
\frac{\lambda}{4!}\Phi^4(\vec{r},t) =
\frac{\lambda}{4}\langle\Phi^2(\vec{r},t)\rangle
\Phi^2(\vec{r},t)+
\left(\frac{\lambda}{4!}\Phi^4(\vec{r},t)-
\frac{\lambda}{4}\langle\Phi^2(\vec{r},t)\rangle\Phi^2(\vec{
r},t)\right)
\label{hart}
\end{equation}
the first term is absorbed in a shift of the mass term
\[m^2(t) \rightarrow
m^2(t)+\frac{\lambda}{2}\langle\Phi^2(t)\rangle \]
(where we used spatial translational invariance). The second
term in
(\ref{hart}) is taken as an interaction with the term
$\langle\Phi^2(t)\rangle\Phi^2(\vec{r},t)$ as a ``mass
counterterm''.
The Hartree
approximation consists of  requiring that the one loop
correction to the two
point Green's functions must be cancelled by the ``mass
counterterm''. This
leads to the self consistent set of equations
\begin{equation}
\langle\Phi^2(t)\rangle  =  \int
\frac{d^3k}{(2\pi)^3}\left(-
iG_k^{<}(t,t)\right) =
\int \frac{d^3k}{(2\pi)^3} \frac{1}{2\omega_{<}(k)}
{\cal{U}}^+_k(t)
{\cal{U}}^-_k(t) \coth[\beta_i\omega_{<}(k)/2] \label{fi2}
\end{equation}
\begin{equation}
\left[\frac{d^2}{dt^2}+\vec{k}^2+m^2(t)+\frac{\lambda}{2}\langle
\Phi^2(t)\rangle\right]
{\cal{U}}^{\pm}_k=0 \label{hartree}
\end{equation}

Before proceeding any further, we must address the fact that
the composite
operator $\langle\Phi^2(\vec{r},t)\rangle$ needs one
subtraction and
multiplicative
renormalization. As usual the subtraction is absorbed in a
renormalization
of the bare mass, and the multiplicative renormalization
into a
renormalization of the coupling constant. We must also point
out that the
Hartree approximation is uncontrolled in this scalar theory;
it becomes
equivalent to the large-N limit in theories in which the
field is in the vector
representation of O(N) (see for example\cite{dolan}).

At this stage our justification
for using this approximation
is based on the fact that it provides a non-perturbative
framework to sum
an infinite series of Feynman diagrams of the cactus type\cite{dolan,chang}.

In principle one may improve on this approximation
 by using
the Hartree propagators in a loop expansion. The cactus-type
diagrams will
still be cancelled by the counterterms (Hartree condition),
but other
diagrams with loops (for example diagrams with multiparticle
thresholds)
may be computed by using the Hartree propagators on the
lines. This approach will have the advantage that the Hartree propagators
will only be unstable for a finite time $ t \leq t_s$. It is not presently
clear to these authors, however, what if any,  would be the expansion
parameter in this case.

It is clear that for $t<0$ there is a self-consistent
solution to the
Hartree equations with equation (\ref{fi2}) and
\begin{eqnarray}
\langle\Phi^2(t)\rangle       & = &  \langle\Phi^2(0^-)\rangle \nonumber \\
{\cal{U}}^{\pm}_k             & = &  \exp[\mp i \omega_{<}(k)] \\
\omega^2_{<}(k)               & = &  \vec{k}^2+m^2_i+\frac{\lambda}{2}+
\langle\Phi^2(0^-)\rangle = \vec{k}^2+m^2_{i,R} \nonumber \\
\end{eqnarray}

where the composite operator has been absorbed in a renormalization of the
initial mass, which is now parametrized as
$m^2_{i,R}=\mu^2_R[(T^2_i/T^2_c)-1]$. For
$t>0$ we subtract the composite operator at $t=0$
absorbing the subtraction
into a renormalization of $m^2_f$ which we now parametrize
as $m^2_{f,R}=
\mu^2_R[1-(T^2_f/T^2_c)]$. We should point out that this
choice of
parametrization only represents a choice of the bare
parameters, which can
always be chosen to satisfy this condition. The logarithmic
multiplicative
divergence of the composite operator will be absorbed in a
coupling constant
renormalization consistent with the Hartree approximation\cite{note},
however, for the purpose of understanding the dynamics of growth of
instabilities associated with the long-wavelength fluctuations,
we will not need to specify this procedure. After
this subtraction
procedure, the Hartree equations read
\begin{equation}
[\langle\Phi^2(t)\rangle-\langle\Phi^2(0)\rangle]  =
\int \frac{d^3k}{(2\pi)^3} \frac{1}{2\omega_{<}(k)}
[{\cal{U}}^+_k(t)
{\cal{U}}^-_k(t)-1] \coth[\beta_i\omega_{<}(k)/2]
\label{subfi2}
\end{equation}
\begin{equation}
\left[\frac{d^2}{dt^2}+\vec{k}^2+m^2_R(t)+\frac{\lambda_R}{2}
\left(\langle\Phi^2(t)\rangle-\langle\Phi^2(0)\rangle\right)
\right]
{\cal{U}}^{\pm}_k(t)=0 \label{subhartree}
\end{equation}
\begin{equation}
m^2_R(t)= \mu^2_R \left[\frac{T^2_i}{T^2_c}-1\right]
\Theta(-t)
- \mu^2_R \left[1-\frac{T^2_f}{T^2_c}\right] \Theta(t)
\end{equation}
with $T_i > T_c$ and $T_f \ll T_c$.
With the self-consistent solution and boundary condition for
$t<0$
\begin{eqnarray}
[\langle\Phi^2(t<0)\rangle-\langle\Phi^2(0)\rangle]  & = & 0
\label{bcfi2} \\
{\cal{U}}^{\pm}_k(t<0)       & = & \exp[\mp i
\omega_{<}(k)t]
\label{bcmodes}\\
\omega_{<}(k)                & = & \sqrt{\vec{k}^2+m^2_{iR}}
\end{eqnarray}

This set of Hartree equations is extremely complicated
to be solved exactly.
However it has the correct physics in it. Consider the
equations for $t>0$,
at very early times, when (the renormalized) $\langle\Phi^2(t)\rangle-
\langle\Phi^2(0)\rangle \approx 0$
the mode functions are the same as in the zeroth order approximation,
and the unstable modes grow exponentially. By computing the expression
(\ref{subfi2}) self-consistently  with
these zero-order unstable modes, we see that the fluctuation
operator begins to grow exponentially.

As $(\langle\Phi^2(t)\rangle-\langle\Phi^2(0)\rangle)$ grows
larger,
its contribution to the Hartree equation tends to balance
the negative
mass term, thus weakening the unstabilities, so that only longer
wavelengths can become
unstable. Even for very weak coupling constants,
the exponentially
growing modes make the Hartree term in the equation of
motion for the mode
functions become large and compensate for the negative mass
term.
Thus when

\[\frac{\lambda_R}{2m^2_{f,R}}\left(\langle\Phi^2(t)\rangle-
\langle\Phi^2(0)\rangle\right)
\approx 1 \]
the instabilities
shut-off, this equality determines the ``spinodal time'' $t_s$.
The modes will still continue to grow further
after this point
because the time derivatives are fairly (exponentially)
large, but eventually
the growth will slow-down when fluctuations sample deep
inside the stable  region.

After the subtraction, and multiplicative renormalization
(absorbed in a
coupling constant renormalization), the composite operator
is finite. The
stable mode functions will make a {\it perturbative}
contribution to the
fluctuation which will be always bounded in time.  The most
important contribution will be that of the {\it unstable
modes}. These will grow
exponentially at early times and their effect will dominate
the dynamics of
growth and formation of correlated domains. The full set of
Hartree equations
is extremely difficult to solve, even numerically, so we
will restrict
ourselves to account {\it only} for the unstable modes. From
the above
discussion it should be clear that these are the only
relevant modes for the
dynamics of  formation and growth  of domains, whereas the
stable modes,  will
always contribute perturbatively for weak coupling after renormalization.

Introducing the dimensionless ratios (\ref{dimensionless})
in terms of
$m_{f,R}\; ; \; m_{i,R}$, (all momenta are now expressed in
units of
$m_{f,R}$), dividing (\ref{subhartree}) by $m_{f,R}^2$,
using the high temperature
approximation $\coth[\beta_i\omega_{<}(k)/2] \approx
2T_i/\omega_{<}(k)$
for the unstable modes, and expressing the critical
temperature as
$T^2_c=24 \mu_R/\lambda_R$, the set of Hartree equations
(\ref{subfi2},
\ref{subhartree}) become the following integro-differential
equation for
the mode functions for $t>0$
\begin{equation}
\left[\frac{d^2}{d\tau^2}+q^2-1+g\int^1_0 dp
\left\{\frac{p^2}{p^2+L^2_R}
[{\cal{U}}^+_p(t){\cal{U}}^-_p(t)-1]\right\}
\right]{\cal{U}}^{\pm}_q(t)=0
\label{finalhartree}
\end{equation}
with
\begin{eqnarray}
{\cal{U}}^{\pm}_q(t<0) & = & \exp[\mp i \omega_{<}(q)t]
\label{bounconhart} \\
\omega_{<}(q)          & = & \sqrt{q^2+L^2_R} \label{frequ}
\\
L^2_R                  & = & \frac{m^2_{i,R}}{m^2_{f,R}} =
\frac{[T_i^2-T_c^2]}{[T^2_c-T_f^2]}
\\
g                      & = & \frac{\sqrt{24\lambda_R}}{4\pi^2}
\frac{T_i}{[T^2_c-T^2_f]^{\frac{1}{2}}}
\label{effectivecoupling}
\end{eqnarray}
The effective coupling (\ref{effectivecoupling}) reflects
the enhancement of
quantum fluctuations by high temperature effects; for
$T_f/T_c \approx 0$,
and for couplings as weak as $\lambda_R \approx 10^{-12}$,
$g \approx 10^{-7} (T_i/T_c)$.

The equations (\ref{finalhartree}) may now be integrated
numerically for the
mode functions; once we find these, we can then compute the
contribution of the unstable modes
 to the subtracted
correlation function equivalent to  (\ref{integral})
\begin{eqnarray}
{\cal{D}}^{(HF)}(x,\tau)    & = &
\frac{\lambda_R}{6 m_{f,R}^2} \left[\langle \Phi(\vec{r},t)
\Phi(\vec{0},t)\rangle-
\langle\Phi(\vec{r},0)\Phi(\vec{0},0)\rangle\right]
\label{hartreecorr1} \\
3{\cal{D}}^{(HF)}(x,\tau)   & = & g\int_0^1dp
\left(\frac{p^2}{p^2+L^2_R}
\right)\frac{\sin(px)}{(px)}\left[{\cal{U}}^+_p(t){\cal{U}}^
-_p(t)-1\right]
\label{hartreecorr2}
\end{eqnarray}
In figure (4) we show

\[ \frac{\lambda_R}{2m_{f,R}^2}(\langle\Phi^2(\tau)\rangle -
\langle\Phi^2(0)\rangle)=
3({\cal{D}}^{HF}(0,\tau)- {\cal{D}}^{HF}(0,0)) \]
(solid line) and
also for comparison, its zeroth-order counterpart
$3({\cal{D}}^{(0)}(0,\tau)-{\cal{D}}^{(0)}(0,0))$ (dashed
line)
for $\lambda_R = 10^{-12}\; , \; T_i/T_c=2$.
(Again, this value of the initial temperature does not have any
particular physical significance and was chosen as a
representative).  We clearly see what we expected;
 whereas the zeroth order correlation grows indefinitely,
the Hartree
correlation function is bounded in time and oscillatory. At
$\tau \approx
10.52$ ,  $3({\cal{D}}^{(HF)}(0,\tau)-{\cal{D}}^{(HF)}(0,\tau))
= 1$,
fluctuations are sampling field configurations near the
classical spinodal, fluctuations
 continue to grow, however,  because the
derivatives are still fairly large. However, after this time, the
modes  begin  to probe the stable region in which there is no
exponential growth. At this point
$\frac{\lambda_R}{2m_{f,R}^2}(\langle\Phi^2(\tau)\rangle-
\Phi^2(0)\rangle)$,
becomes small again because of the small coupling $g \approx
10^{-7}$, and the correction term becomes small.  When
it becomes
smaller than one, the instabilities set in again, modes
begin to grow and the process repeats.
This gives rise to an oscillatory behavior around
$\frac{\lambda_R}{2m^2_{f,R}}(\langle\Phi^2(\tau)\rangle-
\Phi^2(0)\rangle=1$ as shown in figure (4).

In figures (5.a-d), we show the structure factors as a
function of
$x$ for $\tau = 6, \; 8, \; 10, \; 12$, both for zero-order
(tree
level) ${\cal{D}}^{(0)}$ (dashed lines) and Hartree
${\cal{D}}^{(HF)}$ (solid lines).
These correlation functions clearly show that correlations
grow in amplitude
 and that the size of the region in which
the fields are
correlated increases with time. Clearly this region may be
interpreted as
a ``domain'', inside which the fields have strong
correlations, and outside
which the fields are uncorrelated.

We see that up to the spinodal time $\tau_s \approx 10.52$
at which
 $\frac{\lambda_R}
{2m_{f,R}}(\langle\Phi^2(\tau_s)\rangle-
\Phi^2(0)\rangle)=1$, the zeroth order
correlation
$3{\cal{D}}^{(0)}(0,\tau)$ is very close to the Hartree
result. In fact at
$\tau_s$, the difference is less than $15\%$  In particular for
these values of the coupling and initial temperature, the zeroth
order correlation function leads to $t_s \approx 10.15$, and we may
use the zeroth order correlations to provide an analytic estimate for
$t_s$, as well as  the form of the correlation functions and the
size of the  domains.

The fact that the zeroth-order
correlation remains very close to the Hartree-corrected
correlations up to
times comparable to the spinodal is clearly a consequence of
the very small coupling.

To illustrate this fact, we show in figures (6,7) the same
correlation functions
but for $\lambda = 0.01 , \; T_i/T_c =2$. Clearly the
stronger coupling makes
the growth of domains much faster and the departure from
tree-level
correlations more dramatic. Thus, it becomes clear that for
strong couplings,
domains will form very rapidly and only grow to sizes of the
order of the
zero temperature correlation length. The phase transition
will occur very
rapidly, and clearly our initial assumption of a rapid
supercooling will be unjustified.
 This situation for strong couplings, of domains
forming very
rapidly to sizes of the order of the zero temperature
correlation length,
is the picture presented by Mazenko and collaborators\cite{mazenko}.
However, for very
weak couplings (consistent with the bounds from density
fluctuations), our
results indicate that the phase transition will proceed very
slowly, domains
will grow for a long time and become fairly large,
with a typical size several times the zero
temperature correlation length. In a sense, this is a self
consistent check
of our initial assumptions on a rapid supercooling in the
case of weak couplings.

Thus, as we argued above, for very weak coupling,
we may use the tree level result to give an
approximate bound to the correlation functions up to times
close to the
spinodal time using the result given by equation (\ref{strucfactau}),
for $T_f \approx 0$.
Thus, we conclude that for large times, and very weakly
coupled theories ($\lambda_R \leq 10^{-12}$) and for initial temperatures
of the order of the critical temperature,
the size of the domains $\xi_D(t)$ will grow typically in time
as

\begin{equation}
 \xi_D(t)       \approx  (8\sqrt{2})^{\frac{1}{2}}
 \xi(0)\sqrt{\frac{t}{\xi(0)}}
\label{sizedomain}
 \end{equation}

with $\xi(0)$ the zero temperature correlation length. The
maximum size of
a domain is approximately determined by the time at which
fluctuations begin
probing the stable region, this is the spinodal time $t_s$ and the
maximum size of the domains is approximately $\xi_D(t_s)$.

An estimate for the spinodal time, is obtained from equation
(\ref{strucfactau}) by the condition $3{\cal{D}}(\tau_s)=1$,
then for weakly
coupled theories and $T_f \approx 0$, we obtain
\begin{equation}
\tau_s = \frac{t_s}{\sqrt{2}\xi(0)} \approx -\ln\left[
\left(\frac{3\lambda}{4\pi^3}\right)^
{\frac{1}{2}}\left(\frac{(\frac{T_i}{2 T_c})^3}{[
\frac{T^2_i}{T^2_c}-
1]}\right)\right]
\end{equation}

It is remarkable that the domain size scales as $\xi_D(t)
\approx t^{\frac{1}{2}}$ just like in classical theories
 of spinodal decomposition,
when the order parameter {\it is not conserved}, as is
the case in the
scalar relativistic field theory under consideration,
but certainly for completely different reasons. At the tree
level, we can
identify this scaling behavior as arising from the
relativistic dispersion
relation, and second order time derivatives in the equations
of motion, a
situation very different from the classical description of
the Allen- Cahn-
Lifshitz\cite{langer,guntonmiguel,allen} theory of spinodal
decomposition based on a time-dependent Landau Ginzburg model.

{\bf Beyond Hartree:}

The Hartree approximation, keeping only the unstable modes
in the self-
consistent equation, clearly cannot be accurate for times
beyond the
spinodal time. When the oscillations in the Hartree solution
begin, the field
fluctuations are probing the stable region. This should
correspond to the
onset of the  ``reheating'' epoch, in which dissipative
effects become
important for processes of particle and entropy production.
Clearly
the Hartree approximation ignores all dissipative processes,
as may be
understood from the fact that this approximation sums the
cactus type
diagrams for which there are no multiparticle thresholds.
Furthermore, in
this region, the contribution of the stable modes to the
Hartree equation
becomes important for the subsequent evolution beyond the
spinodal point and
clearly will contribute to the ``reheating'' process.
A possible approach to incorporate the contribution of the
stable modes may be that explored by Avan and de
 Vega\cite{avan} in terms of the effective action for the
 composite operator.

Thus, although the Hartree approximation may give a fairly
accurate picture
of the process of domain formation and growth, one must go
beyond this
approximation at times later than the spinodal time, to
incorporate
dissipative  effects and to study the ``reheating'' period.
Clearly, one must also attempt to study the possibility of
``percolation of domains''. Furthermore, the Hartree
approximation is essentially a Gaussian approximation, as
the wave-functional (or in this case the functional density
matrix) is Gaussian with kernels that are obtained self-
consistently. The wave-functional must include non-gaussian
correlations that will account for the corrections to the
Hartree approximation and will be important to obtain the long time
behavior for $t > t_s$.

\section{\bf Conclusions and Looking Ahead:}

The motivations of this work were twofold. First we pointed
out that the dynamics of typical phase transitions in weakly
coupled theories must be studied away from thermodynamic
equilibrium, and introduced the methods and techniques of
non-equilibrium quantum statistical mechanics to study
this situation.

Second we studied both analytically and numerically the case
of a strongly supercooled phase transition in which the
system initially in thermal equilibrium at an initial
temperature larger than the critical is cooled down to
temperatures well below the transition temperature. The
motivation here was to model a period of rapid inflation in
a weakly coupled theory and to study the formation and
growth of correlated domains. We indicated that the {\it
dynamics} of the phase transition cannot be studied within
perturbation theory  because of the instabilities that drive
the process of domain formation and growth, that is spinodal
decomposition.

We used a non-perturbative self-consistent Hartree
approximation to study the time evolution of domain growth,
reflected in the equal time two point correlation function
$\langle\Phi(\vec{r},t)\Phi(\vec{0},t)\rangle$.

We conclude that for {\it weakly} coupled theories at long
times (and distances)

\begin{eqnarray}
 \langle\Phi(\vec{r},t)\Phi(\vec{0},t)\rangle
               & \approx &  \frac{\exp[\sqrt{2}t/ \xi(0)]}
 {(\sqrt{2}t/ \xi(0))^{\frac{3}{2}}}
\exp[-{r^2}/{\xi^2_D(t)}]
\frac{\sin(\sqrt{8}r/ \xi_D(t))}{(r/ \xi_D(t))}
\label{domainamp} \\
 \xi_D(t)      & \approx & (8\sqrt{2})^{\frac{1}{2}}
 \xi(0)\sqrt{\frac{t}{\xi(0)}}
\label{domainsiz}
 \end{eqnarray}
with $\xi(0)$ the zero temperature correlation length. The
domains, however, will grow up to a maximum time at which
the fluctuations begin sampling the stable region. This
maximum ``spinodal time'' is approximately given for weakly coupled
theories by

\begin{equation}
 t_s \approx -\sqrt{2}\xi(0)\ln
\left[\left(\frac{3\lambda}{4\pi^3}\right)^
{\frac{1}{2}}\left(\frac{(\frac{T_i}{2 T_c})^3}{[
\frac{T^2_i}{T^2_c}-
1]}\right)\right]
\end{equation}

When the self-couplings are strong, the phase transition proceeds rapidly,
and domains will not have time to grow substantially, and their sizes
will be of the order of the zero temperature correlation length.

In principle the ``sudden approximation'' (quenching) may be relaxed at the
expense of complications, however the formalism presented in this paper
is completely general, once the initial state is specified and
 the boundary conditions for the mode functions
are understood, the time evolution of correlation functions may be
studied numerically.

Clearly, the next step is to study the dynamics of the phase transition in
FRW cosmologies. In this case, there are several physical effects that
will play a very important role in the dynamics. In particular,
the  red-shift of physical wave
vectors, will tend to enhance the instabilities, because more wave vectors
are entering the unstable region as time evolves. On the other hand, the
presence of a horizon, and the  ``friction''
term in the Heisenberg equations of motion, will prevent domains from
growing bigger than the
horizon size. Thus there seems to be a competition between the different
time scales that must be studied carefully to obtain any meaningful
conclusion about formation and growth of correlated domains
in FRW cosmologies.

\acknowledgements

The authors would like to thank R. Holman for discussions and illuminating
comments. D. B. and D.-S. Lee gratefully acknowledge partial support from
N.S.F. through grant: PHY-8921311. D.-S. Lee would like to acknowledge
partial support through a Mellon Fellowship at the University of
 Pittsburgh, A. Singh was supported by DOE Grant DE-AC02-76ER03066 .
D.B. would like to deeply thank R. Willey, D. Jasnow, H. de Vega,
P. Ramond, Y. Oono,
I. Lawrie and A. Weldon for illuminating discussions, suggestions and
criticisms.

\newpage

{\bf Figure Captions:}

\underline{\bf Figure 1:} Contour in complex time plane to
evaluate the  generating
functional for non-equilibrium Green's functions.

\underline{\bf Figure 2.a:} Two vertices and four
propagators generate the
Feynman diagramatic expansion for non-equilibrium Green's
functions.

\underline{\bf Figure 2.b:} Diagrams that contribute up to
one loop to
$\langle\Phi_{\vec{k}}(t)\Phi_{-\vec{k}}(t')\rangle$.

\underline{\bf Figure 2.c:} Two  mass counterterms.

\underline{\bf Figure 3:} Zero and one loop contribution to
the structure
factor. The solid line represents
$3{\cal{D}}^{(0)}(0,\tau)$, the dashed
line represents  $3{\cal{D}}^{(1)}(0,\tau)$.

\underline{\bf Figure 4:} Hartree (solid line) and zero
order (dashed line)
results for
$\frac{\lambda_R}{2m^2_f}(\langle\Phi^2(\tau)\rangle-
\langle\Phi^2(0)\rangle) = {\cal{D}}(0,\tau)$,
for $\lambda=10^{-12}$, $\frac{T_i}{T_c}=2$.

\underline{\bf Figure 5.a:} Scaled correlation functions for
$\tau=6$, as function of $x$,
${\cal{D}}^{(HF)}(x,\tau)$ (solid line), and
${\cal{D}}^{(0)}(x,\tau)$ (dashed line).
$\lambda=10^{-12}$, $\frac{T_i}{T_c}=2$.

\underline{\bf Figure 5.b:} Scaled correlation functions for
$\tau=8$,
as function of $x$,
${\cal{D}}^{(HF)}(x,\tau)$ (solid line), and
${\cal{D}}^{(0)}(x,\tau)$
(dashed line).
$\lambda=10^{-12}$, $\frac{T_i}{T_c}=2$.

\underline{\bf Figure 5.c:} Scaled correlation functions for
$\tau=10$, as function of $x$,
${\cal{D}}^{(HF)}(x,\tau)$ (solid line), and
${\cal{D}}^{(0)}(x,\tau)$ (dashed line).
$\lambda=10^{-12}$, $\frac{T_i}{T_c}=2$.

\underline{\bf Figure 5.d:} Scaled correlation functions for
$\tau=12$, as function of $x$,
${\cal{D}}^{(HF)}(x,\tau)$ (solid line), and
${\cal{D}}^{(0)}(x,\tau)$ (dashed line).
$\lambda=10^{-12}$, $\frac{T_i}{T_c}=2$.

\underline{\bf Figure 6:} Hartree (solid line) and zero
order (dashed line)
results for $\frac{\lambda_R}{2m^2_f}(\langle\Phi^2(\tau)
\rangle-\langle\Phi^2(\tau)\rangle) = 3{\cal{D}}(0,\tau)$,
for $\lambda=0.01$, $\frac{T_i}{T_c}=2$.

\underline{\bf Figure 7:} Scaled correlation functions for
$\tau=4.15$, as function of $x$,
${\cal{D}}^{(HF)}(x,\tau)$ (solid line), and
${\cal{D}}^{(0)}(x,\tau)$ (dashed line).
$\lambda=0.01$, $\frac{T_i}{T_c}=2$.


\begin{references}
\bibitem{abbott} For an introductory review on the subject,
see
for example L. Abbott and S.-Y. Pi, Inflationary Cosmology
(World Scientific, 1986).
\bibitem{guth1} A. H. Guth, Phys. Rev. D15, 347 (1981); A.
H. Guth
and E. J. Weinberg, Nucl. Phys. B212, 321 (1983).
\bibitem{guth2} A. H. Guth, Phys. Rev. D23, 347 (1981).
\bibitem{linde1} A. D. Linde, Phys. Lett. B108, 389 (1982).
\bibitem{steinhardt} A. Albrecht and P. J. Steinhardt, Phys.
Rev.
Lett. 48, 1220 (1982).
\bibitem{linde2} For a review see: A. D. Linde, Rep. Prog.
Phys. 47,
925 (1984); P. J. Steinhardt, Comments Nucl. Part. Phys. 12,
273  (1984).
\bibitem{kirzhnits} D. A. Kirzhnits and A. D. Linde, Phys.
Lett. B42, 471, (1972).
\bibitem{dolan} L. Dolan and R. Jackiw, Phys. Rev. D9, 3320
(1974).
\bibitem{sweinberg} S. Weinberg, Phys. Rev. D9, 3357 (1974).
\bibitem{brandenberger} R. H. Brandenberger, Rev. of Mod.
Phys. 57, 1 (1985); Int. J. Mod. Phys. A, 77 (1987).
\bibitem{kolb}  E. W. Kolb and M. S. Turner, ``The Early
Universe'', Addison
Wesley (Frontiers in Physics) (1990)
\bibitem{linde}  A. Linde, Particle Physics
and Inflationary Cosmology, Harwood Academic Publishers
(1990), and references therein.
\bibitem{weakcoup} S. W. Hawking, Phys. Lett.  B 115, 295 (1982);
A. A. Starobinsky, Phys. Lett.  B 117, 175, (1982); A. H. Guth and
S. Y. Pi, Phys. Rev. Lett.  49, 1110 (1982); J. M. Bardeen, P. J.
Steinhardt, M. S. Turner, Phys. Rev  D 28, 679 (1983).
\bibitem{kapusta} For a thorough review on finite temperature field
theory see: J. Kapusta: ``Finite Temperature Field Theory'' Cambridge
Univ. Press (1989); and references therein.
\bibitem{mazenko} G. F. Mazenko, W. G. Unruh and R. M. Wald,
Phys.  Rev. D31, 273 (1985); G. F. Mazenko , Phys. Rev. Lett.
54, 2163 (1985).
\bibitem{boyveg} D. Boyanovsky and H. J. de Vega, ``Quantum Rolling
down out of equilibrium'', preprint Pitt 92-06 (1992)
 (to be published in Phys.Rev.D.).
\bibitem{ringwald} A. Ringwald, Phys. Rev. D36, 2598 (1987);
Ann. of Phys. (N.Y.) 177, 129 (1987).
\bibitem{kripfganz} J. Kripfganz and H. Perlt, Ann. of Phys.
(N.Y.) 191, 241 (1989).
\bibitem{leutwyler} H. Leutwyler and S. Mallik, Ann. of
Phys. (N.Y.) 205, (1991)
\bibitem{eboli} O. Eboli, R. Jackiw and S-Y. Pi, Phys. Rev.
D37, 3557 (1988)
\bibitem{samiulla} M.Samiullah, O. Eboli and S-Y. Pi, Phys.
Rev. D44, 2335 (1991).
\bibitem{wang} E. W. Kolb and Y. Wang, ``Domain wall formation in
late-time phase transitions'', Fermilab preprint FNAL-Pub-92/33-A
(1992).
\bibitem{langer} J. S. Langer in ``Fluctuations,
Instabilities and Phase Transitions) (T. Riste Ed.) page 19,
Plenum N.Y. (1975). See also: J. S. Langer in ``Solids Far From Equilibrium'',
Ed. C. Godreche, Cambridge Univ. Press (1992), page 297.
\bibitem{guntonmiguel} J. D. Gunton, M. San Miguel and P.S.
Sahni in ``Phase Transitions and Critical Phenomena'' (C.
Domb and J. J. Lebowitz, Eds.) Vol 8, Academic Press, (1983).
\bibitem{guthpi} A. Guth and S-Y. Pi, Phys. Rev. D32, 1899
(1985).
\bibitem{weinbergwu} E. J. Weinberg and A. Wu, Phys. Rev.
D36, 2474 (1987).
\bibitem{niemisemenoff} A. Niemi and G. Semenoff, Ann. of
Phys. (N.Y.) 152, 105 (1984); Nucl. Phys. B [FS10], 181,
(1984).
\bibitem{calzetta} E. Calzetta, Ann. of Phys. (N.Y.) 190, 32
(1989)
\bibitem{calzettahu} E. Calzetta and B. L. Hu, Phys. Rev.
D35, 495
(1987); Phys. Rev. D37, 2878 (1988).
\bibitem{schwinger} J. Schwinger, J. Math. Phys. 2, 407
(1961).
\bibitem{keldysh} L. V. Keldysh, Sov. Phys. JETP 20, 1018
(1965).
\bibitem{mills} R. Mills, ``Propagators for Many Particle
Systems''
(Gordon and Breach, N. Y. 1969).
\bibitem{jordan} R. D. Jordan, Phys. Rev. D33, 444 (1986).
\bibitem{landsman} N. P. Landsman and C. G. van Weert, Phys.
Rep. 145, 141 (1987).
\bibitem{semenoffweiss} G. Semenoff and N. Weiss, Phys. Rev.
D31, 689; 699 (1985).
\bibitem{kobeskowalski} R. L. Kobes and K. L. Kowalski,
Phys. Rev.
D34, 513 (1986); R. L. Kobes, G. W. Semenoff and N. Weiss,
Z. Phys.C 29, 371 (1985).
\bibitem{paz} J. P. Paz, Phys. Rev. D41, 1054 (1990); Phys.
Rev. D42, 529 (1990).
\bibitem{conjecture} We believe that this approximation is somehow
equivalent to the variational approach advocated in
references\cite{eboli,samiulla}. We are currently studying this connection.
\bibitem{lawrie} An alternative approach based on the composite operator
$ \Phi^2 $ has been advanced by: I. D. Lawrie, Nucl. Phys.  B 301, 685
(1988);
 in ``Advances in Phase Transitions and Disorder
Phenomena'',
Eds. G. Busiello, et. al. (World Scientific), 548 (1987),
Nucl. Phys.  B 301, 685 (1988); J. Phys.  A 21: Math. and Gen.,
L823. (1988).
\bibitem{chang} For a treatment of a scalar field theory in the Hartree
approximation see also S.J. Chang,  Phys. Rev.  D 12, 1071 (1975).
\bibitem{note} By using the WKB procedure used in reference\cite{boyveg}
or by following the renormalization procedures of
reference\cite{dolan,chang}, we find that for $t>0$
\begin{displaymath}
\int \frac{d^3k}{(2\pi)^3} \frac{1}{2\omega_{<}(k)}
[{\cal{U}}^+_k(t)
{\cal{U}}^-_k(t)-1] \coth[\beta_i\omega_{<}(k)/2]|_{div} =
\frac{\lambda}{8\pi^2}
[\langle \Phi^2(t) \rangle-\langle\Phi^2(0) \rangle]
\ln(\frac{\Lambda}{\kappa}) \nonumber
\end{displaymath}
the logarithmic divergence will be absorbed in the renormalization of the
coupling constant. We will be cavalier about the renormalization procedure
as we are only interested in the unstable fluctuations which will always
yield to a finite and unambiguos contribution, and will refer to the
renormalized (subtracted and multiplicatively renormalized) composite
operator.
\bibitem{allen} S. M. Allen and J. W. Cahn, Acta Met.
27, 1017,1085 (1976);
I. M. Lifshitz, Sov. Phys. JETP {\bf 15}, 939, (1962)
\bibitem{avan} J. Avan and H. J. de Vega, Phys. Rev. D29, 2891,
2904 (1984)
\end{references}
\end{document}